\documentclass[referee]{raa}           

\usepackage{graphicx,times}
\usepackage{natbib}
\usepackage{amssymb,amsmath}
\bibpunct{(}{)}{;}{a}{}{,}

\usepackage[pagebackref=true]{hyperref}
\usepackage{enumerate}

\usepackage{color}

\begin{document}

   \title{Astrometric Reduction of Saturnian Satellites with Cassini-ISS Images Degraded by Trailed Stars}

 \volnopage{ {\bf 20XX} Vol.\ {\bf X} No. {\bf XX}, 000--000}
   \setcounter{page}{1}

   \author{Qing-Feng Zhang
   \inst{1,2}, Meng-Qi Liu\inst{1}, Yan Li\inst{1,2} \thanks{Corresponding author: tyanei@jnu.edu.cn}, Lin-Peng Wu\inst{1}, Zhi-Qiang Wang\inst{1}, Li-Sha Zhu\inst{3}, Zhan Li\inst{1,2}
   }

   \institute{ Department of Computer Science, Jinan University, Guangzhou 510632, P. R. China\\
        \and
             Sino-French Joint Laboratory for Astrometry, Dynamics and Space Science, Jinan University, Guangzhou 510632, P. R. China\\
    \and
    Department of Financial Technology, School of Management, Guangzhou Xinhua University, Guangzhou 510520, P. R. China\\
\vs \no
   {\small Received 20XX Month Day; accepted 20XX Month Day}
}

\abstract{Imaging Science Subsystem (ISS) mounted on the Cassini spacecraft has taken a lot of images, which provides an important source of high-precision astrometry of some planets and satellites. However, some of these images are degraded by trailed stars. Previously, these degraded images cannot be used for astrometry. In this paper, a new method is proposed to detect and compute the centers of these trailed stars automatically. The method is then performed on the astrometry of ISS images with trailed stars. Finally, we provided 658 astrometric positions between 2004 and 2017 of several satellites that include Enceladus, Dione, Tethys, Mimas and Rhea. Compared with the JPL ephemeris SAT427,  the mean residuals of these measurements are 0.11 km and 0.26 km in right ascension and declination, respectively. Their standard deviations are 1.08 km and 1.37 km, respectively. The results show that the  proposed method  performs astrometric measurements of Cassini ISS images with trailed stars effectively. 
}

   \authorrunning{Q.-F. Zhang et al. }            
   \titlerunning{Astrometry of Cassini-ISS Images with streaks}  
   \maketitle

%
\section{Introduction}           
\label{sect:intro}

Imaging Science Subsystem (ISS) is a piece of optic equipment mounted on the Cassini spacecraft \citep{Porco2004Cassini}. During the entire Cassini tour, it took more than 400,000 images. These images have provided a lot of astrometric data on the satellites of Saturn and Jupiter. For instance, \citet{cooper2006cassini} obtained the astrometric data of inner  Jovian satellites from the early ISS images.
\citet{tajeddine2013astrometric,tajeddine2015cassini} and \citet{Cooper2014Cassini} reduced a part of ISS images of several main Saturnian satellites. \citet{2018MNRAS.481...98Z,zhang2021comparison,Zhang2022Complementary} measured some ISS images of several Saturnian satellites. These data have been used in the studies of the dynamics of planetary systems \citep{Cooper2015Inner,Desmars2013Phoebe,Lainey2017new,Lainey2020Resonance}. All these researches demonstrate the importance of the astrometry of ISS images.

\begin{figure*} [t!]
	\centering
	\subfloat[\label{fig:a}]{
		\includegraphics[width=0.5\textwidth]{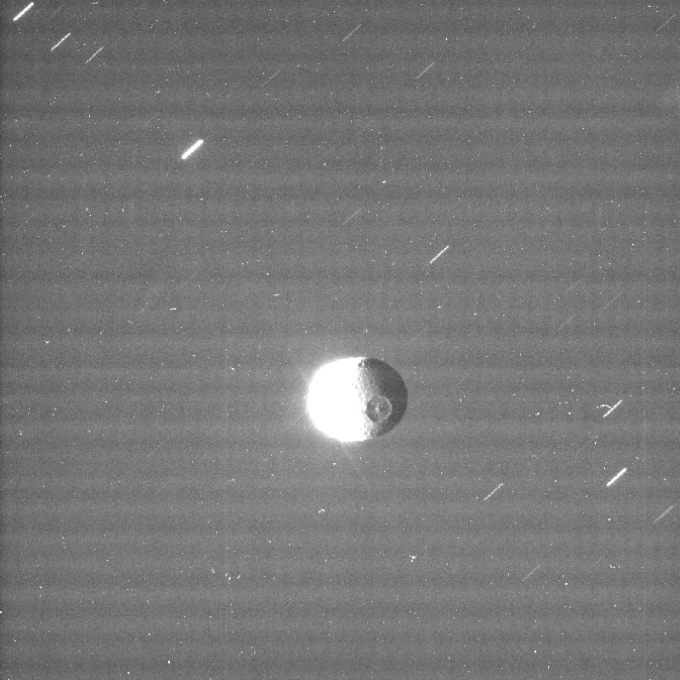}}
	\subfloat[\label{fig:b}]{
		\includegraphics[width=0.5\textwidth]{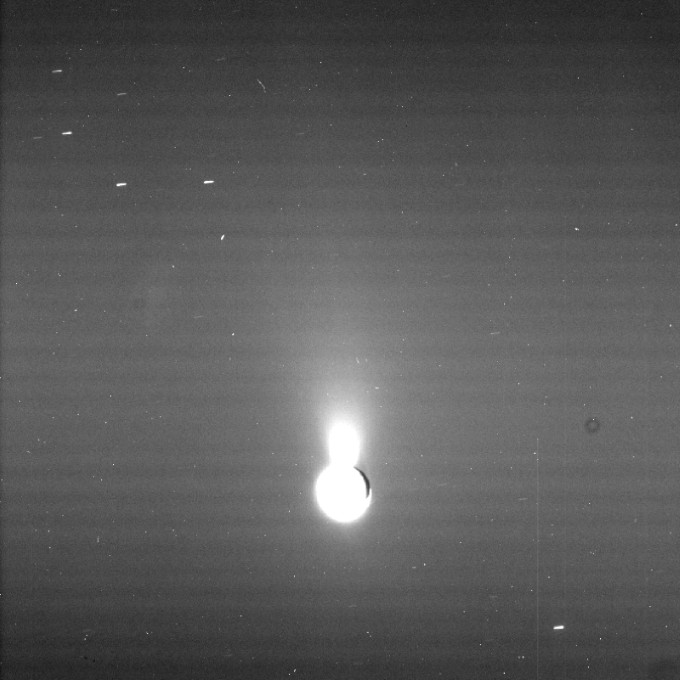}}
	\\
        \centering
	\subfloat[\label{fig:c}]{
		\includegraphics[width=0.5\textwidth]{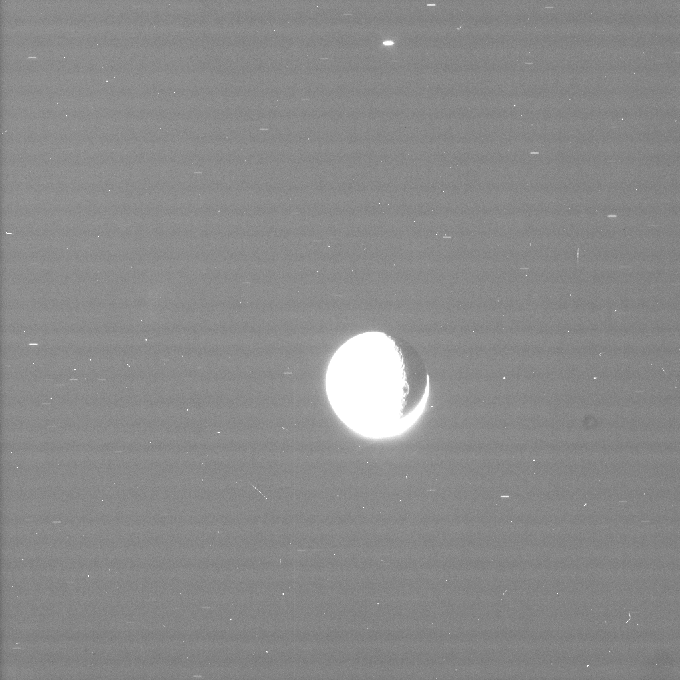} }
	\subfloat[\label{fig:d}]{
		\includegraphics[width=0.5\textwidth]{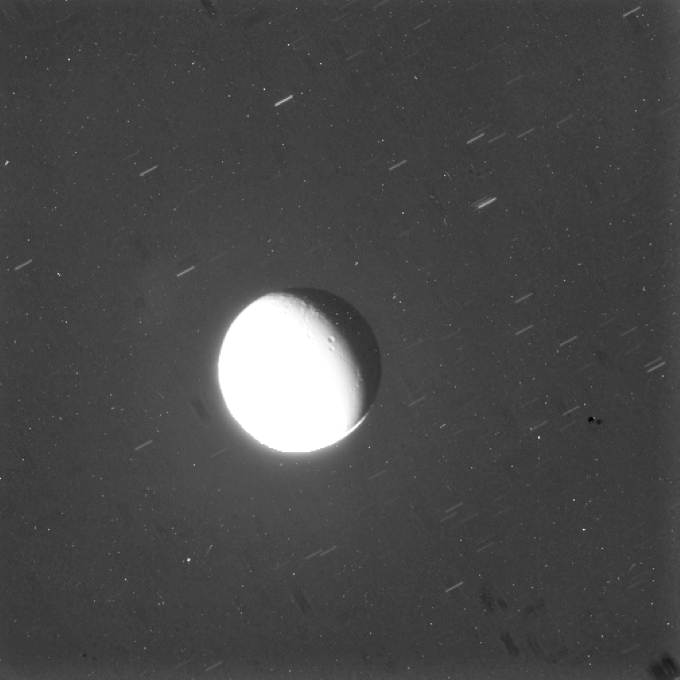}} 
     \caption{  Some examples of ISS images in which all the stars are streaks. The images are enhanced by applying contrast stretching for visibility. (a)  An image of Mimas (Image ID: N1743596513),  (b) An image of  Enceladus (Image ID: N1872048342), (c) An image of Rhea (Image ID: N1516373834), (d) An image of Dione (Image ID: N1880313867).  } 
   \label{figure1}
\end{figure*}

In 2018, a standard and versatile tool for the astrometry of ISS images, Caviar, has been released to the community by  \citet{cooper2018caviar}. Caviar can reduce normal ISS images. However,
it cannot cope with the degraded images with trailed stars because of the inability of detecting streaks and obtaining their centers. In astrometry, these streaks are expected to be used as reference stars to correct  camera pointing. If they cannot be  analyzed, the measurement of the image will fail. Otherwise, these degraded images will be valuable for astrometry and can provide more astrometric positions of celestial objects. Figure \ref{figure1} shows four example images with streaks. They are the ISS images of Mimas, Enceladus, Rhea and Dione, respectively. The streaks were caused by long exposure time or improper tracking. 
\par In these days, detecting and centering streaks in astronomical images has been studied by various researchers. \citet{laas2009new} proposed an algorithm  based on mathematical morphology for detecting the trajectory of space debris and developed the software TAROT that can fully automated extracting source positions of objects in geostationary transfer orbit. It is dedicated to  observations with two small TAROT telescopes.
\citet{sun2013innovative} presented another mathematical morphology method to detect streaks of space debris and obtain its center. As a result, the astrometric precision of  space debris is improved.  However, both of  the above two algorithms need  predict the approximate direction or length of the object to define a structural element. It is not suitable for processing ISS images because the direction and length of streaks are unknown in ISS images. \citet{virtanen2016streak} gave an automated streak detection and processing pipeline for the ESA-funded StreakDet (streak detection and astrometric reduction) activity. Although its detection sensitivity for bright objects or objects longer than 100 pixels reaches $90\%$,  the sensitivity for shorter ones is only $63\%$. Since the length of streaks in ISS images are mostly shorter than 100 pixels, the method is not suitable for ISS images. \citet{sease2017automatic}  offered a method to automatically locate two endpoints of one streak according to corner metrics. But, the centering method of streak was not provided, and the endpoints locating method was tested only on the simulated images. In addition, there are some famous processing systems that contain the function of detecting and centering streaks. For example, Pan-STARRS Moving Object Processing System (MOPS) \citep{Denneau2013Pan} and  TRailed Image Photometry in Python (TRIPPy) \citep{Fraser2016TRIPPy}. All these methods and pipelines are not suitable for the astrometry of ISS images because most of them are dedicated to some specific equipment,  or to a specific project, or aiming at  photometry.

In this paper, a method that can automatically detect and center all streaks in an ISS image is first proposed,  the method is then applied to the astrometry of ISS images with trailed stars of several main Saturnian satellites. The remaining of this paper is organized as follows. An introduction to our streak centering method is presented in Section \ref{sect:section2}. Section \ref{sect:section3} describes the astrometry of several moons of Saturn.  Section \ref{sect:section4} provides a discussion about our method and measurements. The conclusions are given in Section \ref{sect:section5}.
\section{streak centering}
\label{sect:section2}
In the astrometry of ISS images with trailed stars, the key problem is to determine the accurate centers of streaks. There are three main steps in our method to solve the problem, including searching peak points, finding streaks and centering streaks. Details are given below.

\subsection{ Step 1: Searching peak points}
\citet{stetson1987daophot} published a classical package of photometry, DAOPHOT, which includes a routine FIND. FIND routine is very powerful in searching stars in an image. Its principle is to find stars as the points with local maxima in an image $H$ that is generated by convolving an original image $F$ with a truncated Gaussian function, and obtain the star centers by fitting a bivariate Gaussian function to the intensity distribution over their neighborhood in $F$. If the stars are point-like, FIND routine can identify them and output their centers accurately. However, FIND cannot handle the trailed stars correctly. That is, a trailed star may not be found, or may be recognized as one or more stars. Even when the streak is recognized as one star, its center cannot be located correctly. 
\par In this step, the aim is to locate at least one peak point in each streak in image $H$. The peak points will be the input of the next step. Since the streaks of stars in the ISS images are longer than normal point sources, to detect peak points successively the FIND parameter settings should be optimized as follows. The ROUND parameter in FIND is set to ${\pm 2}$, such that stars with any long-width ratio will be accepted. In order to reduce the number of false detections, the Hmin parameter should be increased properly. Based on experimental experiences, we set Hmin to  $25+bakground(H)$, which  means that the H value of Peak points should be at least 25 higher than that of the background.  Readers can refer to \cite{stetson1987daophot} for detail of the parameters ROUND and Hmin. In one word, we use the DAOPHOT FIND routine with proper parameters to find peak points. 
\par Figure \ref{figure2} shows the results of each step of our method. The Left image illustrates the result of the step of searching peak points. It  results from the FIND routine with optimized parameters. In the image, the purple boxes represent detected stars whose centers are at the center positions of these boxes. The content in the small red box is enlarged and shown at the right top of the image. It can be seen from the image that at least one peak point is recognized for each trailed star with optimized parameters

 \subsection{Step 2: Finding streaks}
With at least one peak point in each streak,  the region growing algorithm is used to obtain the streak regions.  The algorithm can find all pixels in the connected neighbors of a seed pixel (i.e. the peak pixel).  For the details of region growing, see also \cite{2002dip..book.....G}. The whole process of finding the streak region consists of six secondary steps as follows.

\begin{enumerate}[(1)]
\item Set a threshold $d$ and  the size of a square local image $n$. $d$ will be used to determine whether the region is a streak or not. $n$ will define the size of the square local image of a peak point.
\item  Given a peak point $P$, find out a local image $SubImg$ that is centered at $P$ and  has a size  $n \times n$.  If $P$ is near the boundary of the image, center $SubImg$ as close to $P$ as possible. 
\item  For the local image $SubImg$, compute the mean ($m$) and standard deviation ($\sigma$) of all pixel intensities after applying 3$\sigma$ clipping criteria. Set intensity threshold $ T=m+2\sigma $.
\item The region growing algorithm is performed to expand the region around point  $P$  and then obtain the expanded region $R$ associated with  $P$. In the region,  all pixels are connected and their intensity is greater than $T$.  The IDL function REGIONGROW is called to finish it. 
\item For each peak point $P_i$ , repeat steps 2-4 to obtain its expanded region  $R_i$ . To avoid overlapping regions, we restrict a pixel to belonging to only one region. Generally,  the restriction  makes sure that one streak has only one expanded region even if it has several peak points. 
\item Determine which expanded regions are streaks. Compute the   maximum length of the region $R_i$ in $x$ , $y$ directions that are denoted by  $L_x$ and $L_y$,  respectively. If $max(L_x,L_y) > d$ , $R_i$ is a streak region and saved as $SR_i$. Perform the same operation on every expanded region to get all streaks.

\end{enumerate}

Through the above six steps,  streak regions are determined. The parameters $d, n$ and $T$  play  important roles in the whole procedure of finding streak regions. $d$ is used to determine whether the region is a streak or not. $n$ defines the size of the square local image $SubImg$ of a peak point, and indirectly determines the intensity threshold $T$ that affects the growth of the region from one peak point.  Based on the properties of ISS images, the parameters are set as  $d=6$, $n=128$ and $T=m+2\sigma$. All these values work well for ISS images. In other applications, these parameters should be adjusted based on the properties of images used.
\par
\begin{figure} 
   \centering
   \includegraphics[width=0.3\textwidth]{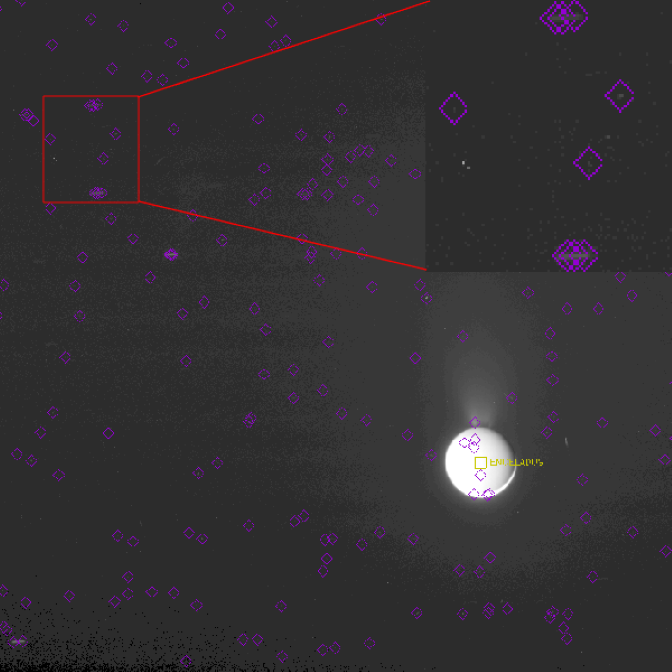}\,\includegraphics[width=0.3\textwidth]{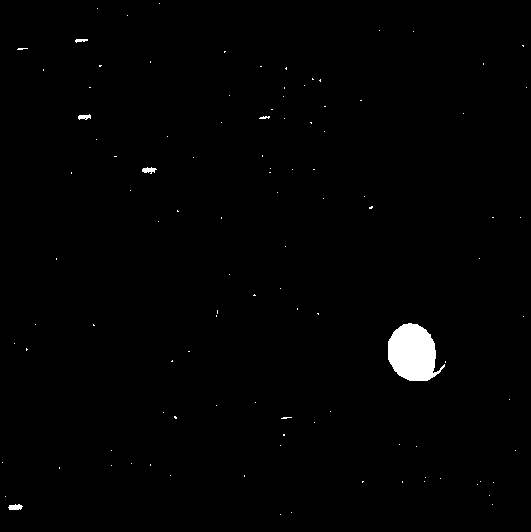}\,\includegraphics[width=0.3\textwidth]{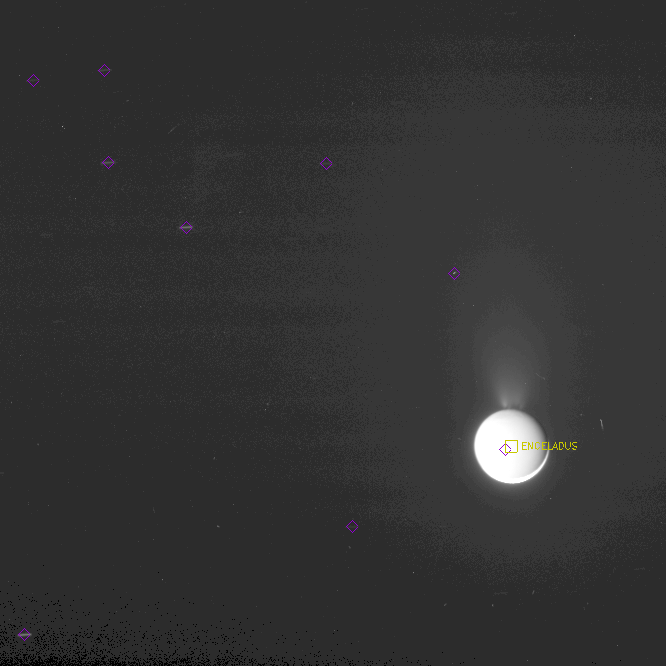}
   \caption{The results of the three steps with a trailed star degraded image. Left: Searched peak points. Mid: Detected streaks. Right: Streaks and their centers.
   } 
   \label{figure2}
   \end{figure}

 The middle image in Figure \ref{figure2}  shows the streaks detected from the left image with our method. It is a binary image, white regions are detected streaks. Comparing with the left image, it can be seen that the number of false detected streaks reduced significantly.

\subsection{Step 3: Centering streaks}
After finding out all streak regions, an improved modified moment method is used to compute the centers of all streaks. Similar to the method described by \cite{zhang2021comparison},  we developed a  modified moment method to locate the centers of streaks. It includes three secondary steps as follows.
\begin{enumerate}[(1)]
\item Obtain a complete streak region $FR_i$. The complete region  should contain both foreground and background areas of a streak. We assume that it is a rectangular area. Since in most cases $SR_i$, which approximates the real foreground area, is an irregular area. To obtain the complete region, we first generate the bounding box of $SR_i$, and then expand it outward by more $k$ pixels. In our processing, the $k$  is 4.
\item Update the intensities of all pixels in $FR_i$. In the approximate background area  $FR_i-SR_i$, a $2\sigma$ clipping method is used to remove outliers to  get all background pixels of a peak point. With the mean ($m$) and standard deviation ($\sigma$) of all background pixel intensities, another threshold  $B=m+2\sigma$ is obtained. All pixels with an intensity no greater than $B$ in $FR_i$  will be changed to zero.  
\item Compute the center of a streak. The modified moment formulation is applied to the region  $FR_i$ to get the streak center.
\end{enumerate}
The secondary steps 1-3 are repeated for each streak region to locate the centers. The right image in Figure \ref{figure2} shows the results from centering streak of the left one. each purple box indicates one streak center. Although there are still false detected streaks, they will not affect the final result because they cannot be matched with catalogue stars.
\par The above 3 main steps comprise the proposed automatic detecting and centering method for streaks in ISS images. In section \ref{sect:section4} a discussion about the method will be given.

\section{Data reduction}  
\label{sect:section3}
We collected 710 ISS images with trailed stars of several moons of Saturn that include Enceladus, Dione, Mimas, Rhea and Tethys. Sifting and discarding the incomplete or contaminated images, a total of 658 ISS images remain.

The Caviar software package \citep{cooper2018caviar} is used to perform the astrometry of these images. In order to enable Caviar to process images with streaks, we integrated our automatic streak centering method into it. 
The whole reduction involves two key operations:  pointing correction and measurement of target centers. In point correction, the catalogue stars were matched with image stars to correct the nominated pointing of ISS camera. The center positions of all image stars were determined by our streak centering method proposed in the paper, and corrected with the geometric distortion model given in \citep{owen2003cassini}. The catalogue stars are extracted from Gaia Early Data Release 3 \citep{2016A&A...595A...1G,brown2021gaia}.  In the measurement of the target's center,  the limb-fitting method (\cite{tajeddine2013astrometric}) is used to get the center  because targets are resolved in all ISS images.   
Finally, we reduced 658 ISS images, which includes 539 images of Enceladus from 2012 to 2017,  57 images of Dione, 22 images of Mimas, 14 images of Rhea and 26 images of Tethys. A sample of the results is given in table \ref{table1}. Each row shows one observation of a target. Column 1 is the ID of the reduced image. Column 2 is the observation time of the image, which is the middle time of the exposure of the image. Column 3 is the target name. Columns $\alpha,\delta$ are the right ascension  and declination of target in ICRF (International Celestial Reference Frame) centered at Cassini. Columns $\alpha_c,\delta_c$ and Twist are the camera's pointing in ICRF centered at Cassini. The corresponding image coordinates are given in columns Sample and Line. The full table \ref{table1} is available at CDS via anonymous ftp to cdsarc.u-strasbg.fr 
(130.79.128.5) or via \url{ https://cdsarc.unistra.fr/viz-bin/cat/J/other/RAA}.

\begin{table}
\bc
\begin{minipage}[]{150mm}
\caption[]{A sample of the results. Column 1 is the Cassini ISS image ID. Column 2 is the date and exposure mid-time of the image (UTC). Columns  \(\alpha\) and \(\delta\) are the right ascension and declination in ICRF centered at Cassini for target, while $\alpha_c, \delta_c$, and Twist refer to the right ascension, declination, and twist angle of the camera’s pointing in the same coordinate system. The columns of Sample and Line are the observed positions in the image.  The full table is available from the CDS. The origin of the sample, line coordinate system is at the top left of the image, and line increasing downwards and sample to the right. All the angle variables are given in degrees.\label{table1}}\end{minipage}
\setlength{\tabcolsep}{2.5pt}
\small
\scalebox{0.8}{
\begin{tabular}{cccccccccc}
\hline
Image ID & Mid-time(UTC)         & Body & $\alpha$           &  $\delta$          & $\alpha_c$          & $\delta_c$          & Twist & Sample   & Line     \\
                            & (UTC)                 &                       & (degrees)   & (degrees)   & (degrees)   & (degrees)   & (degrees)   & (pixels) & (pixels) \\ \hline
N1828232894                 & 2015-342T01:24:23.436 & ENCELADUS             & 82.0837699  & -4.1726995  & 82.0837677  & -4.1727152  & 3.4236956   & 504.65   & 892.14   \\
N1516373834                 & 2006-019T14:27:36.382 & RHEA                  & 319.2427990 & -1.2519495  & 319.2427505 & -1.2518827  & 173.7164582 & 566.94   & 574.44   \\
N1532687765                 & 2006-208T10:04:43.381 & DIONE                 & 279.7775340 & -13.9993040 & 279.7769920 & -13.9992617 & 264.2913532 & 508.13   & 516.82   \\
N1532778115                 & 2006-209T11:10:31.351 & MIMAS                 & 288.0337287 & -11.9340958 & 288.0336926 & -11.9341869 & 263.9209222 & 496.20   & 521.16   \\
N1525531204                 & 2006-125T14:09:25.071 & TETHYS                & 324.9302203 & -1.9154561  & 324.9302255 & -1.9154309  & 199.4802039 & 561.18   & 579.10   \\ \hline
\end{tabular}}
\ec
\end{table}

\section{Discussion}
\label{sect:section4}
\begin{figure} 
   \centering
   \includegraphics[width=0.5\textwidth]{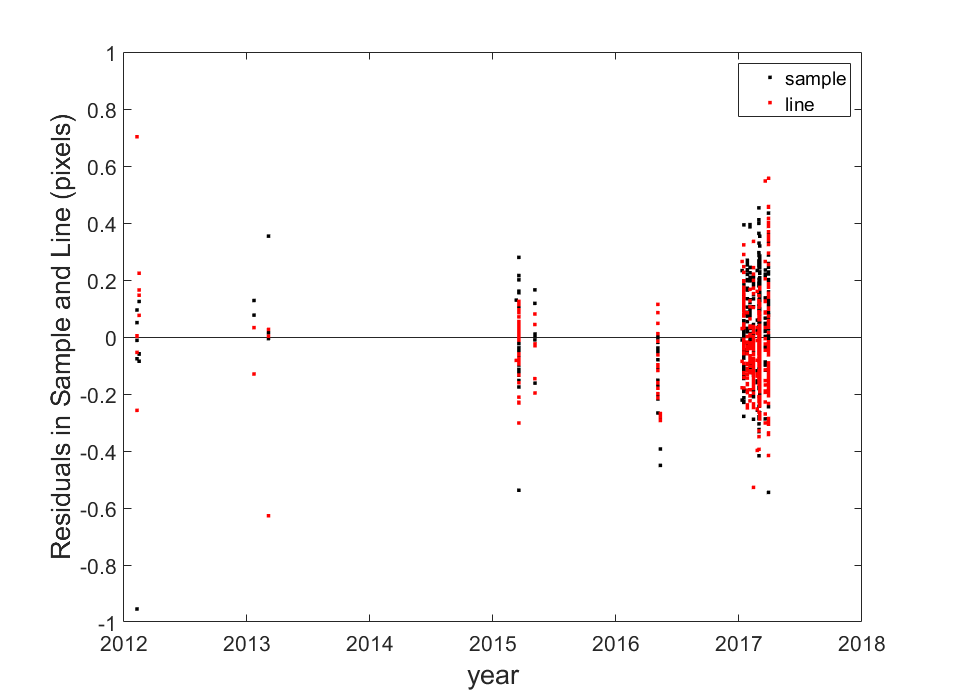}\,\includegraphics[width=0.5\textwidth]{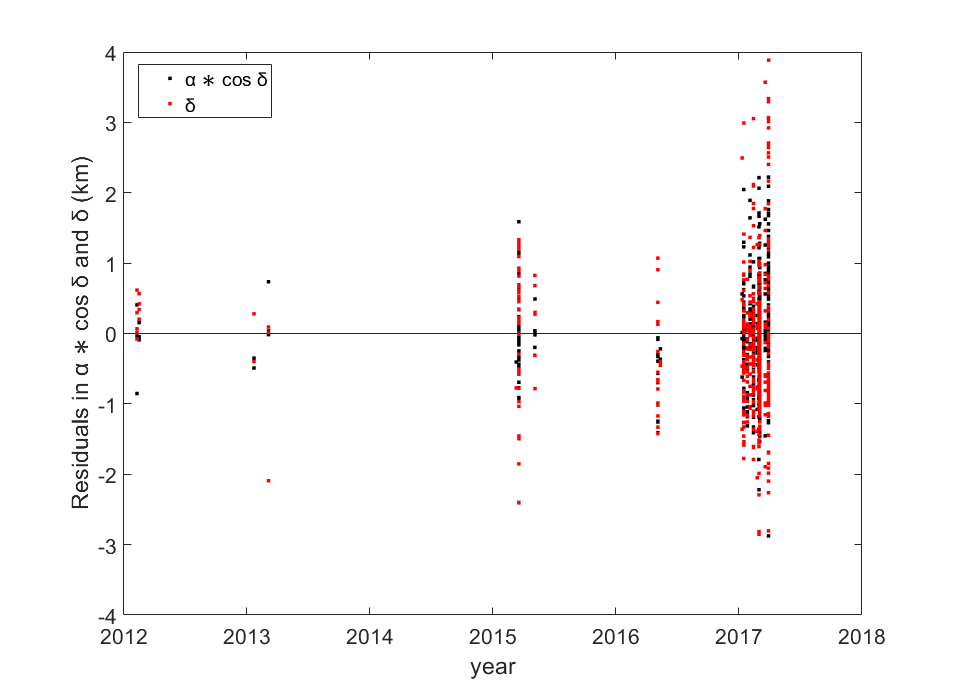}
   \caption{The residuals of Enceladus relative to JPL  ephemeris SAT427 in  Sample and Line (in pixels), and in  \(\alpha*\cos{\delta}\) and \(\delta\) directions (in km)} 
   \label{figure3}
   \end{figure}

\begin{figure} 
   \centering
   \includegraphics[width=0.5\textwidth]{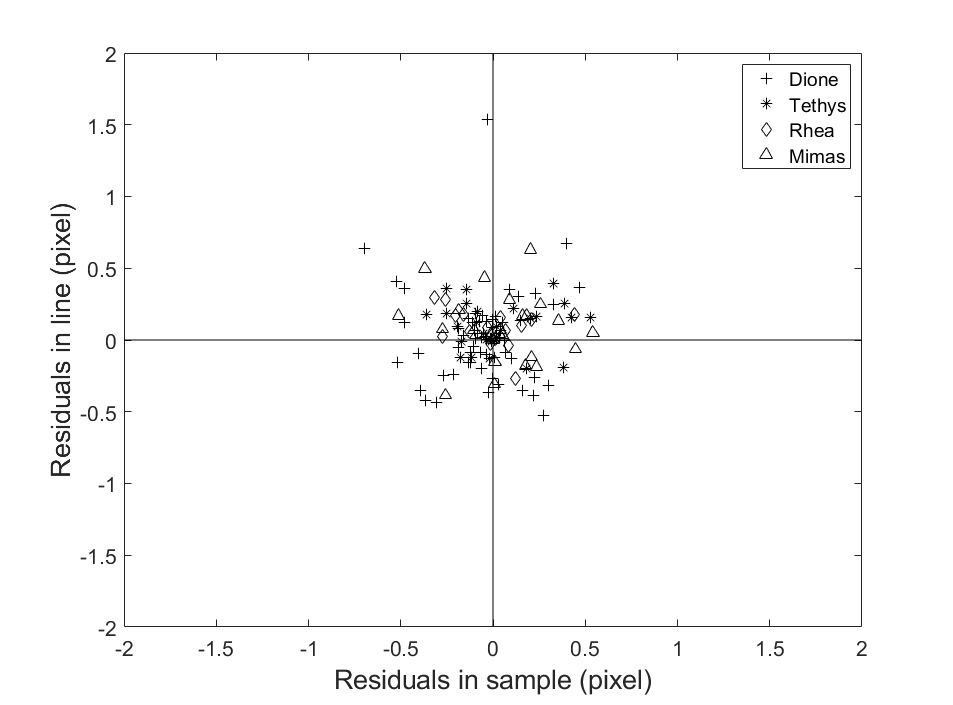}\,\includegraphics[width=0.5\textwidth]{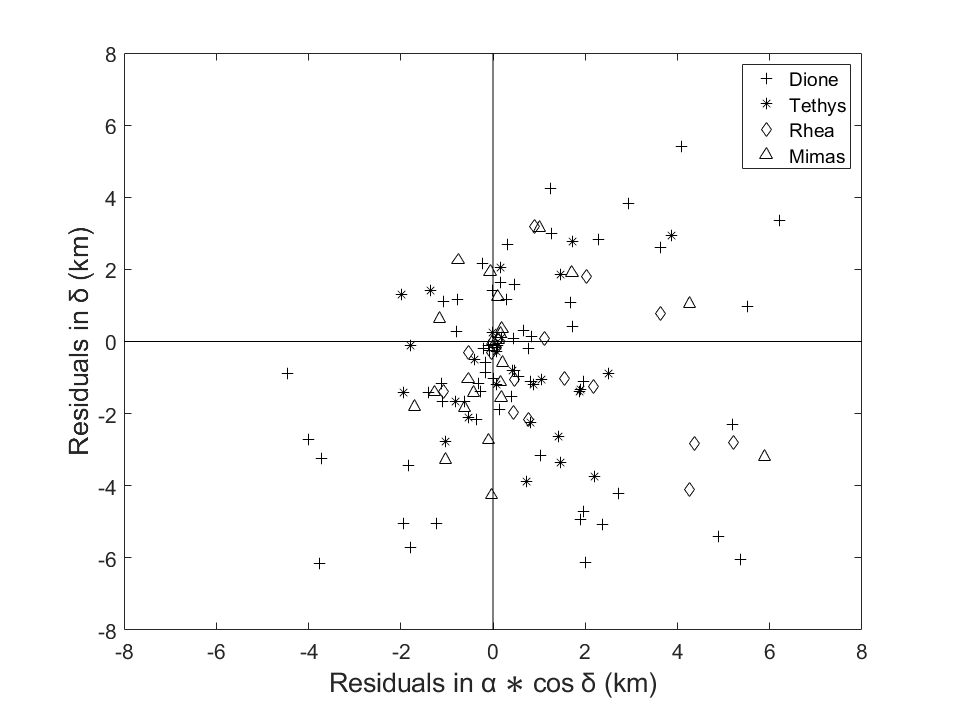}
   \caption{The residuals of Mimas, Dione, Rhea, and Tethys relative to JPL  ephemeris SAT427 in  Sample and Line directions (in pixels), and in  \(\alpha*\cos{\delta}\) and \(\delta\) directions (in km)} 
   \label{figure4}
   \end{figure}

To evaluate our measurement, we compared our results with the JPL ephemeris SAT427 (for detail see also \url{ https://naif.jpl.nasa.gov/pub/naif/generic_kernels/spk/satellites/sat427.cmt})  to get the residuals of all these observations. The residuals of Enceladus in sample and line are displayed in the left panel in Figure \ref{figure3}, and  right ascension ($\alpha$) and declination ($\delta$) are displayed in the right panel in Figure \ref{figure3}.  Figure \ref{figure4} shows the same residuals of other Saturnian satellites, i.e. Mimas, Dione, Rhea and Tethys.

\par The mean and standard deviations (SD) of these residuals are given in table \ref{table2}. It shows that the SD in sample and line of Enceladus are 0.16 pixels and 0.15 pixels, respectively. In terms of distance, the SD in right ascension and declination are  0.69 km and 0.98 km, respectively. For the other four targets: Mimas, Dione, Tethys and Rhea, their counterparts are 0.24 pixels, 0.27 pixels, 2.02 km and 2.36 km, respectively.
\par Table \ref{table2} also shows that the result of Enceladus is more precise than that of the other satellites. There are two reasons. Firstly, the exposure times for Enceladus are generally between 5.6 s and 8.2 s while for other targets are between 10 s and 46 s. The longer exposure time may increase image noise and thus reduce the accuracy of position measurements.  Secondly, the observation conditions of other satellites are more variable than those of Enceladus, which decrease the precision of the measurement.
\par  The results show that our automatic streak centering method works well. As stated in section \ref{sect:section2}, there are a few parameters affecting the method. On one hand, these parameters enhance the flexibility  and application range of our method. On the other hand, these parameters should be set carefully because the improper parameter setting may cause the failure of the method. Our parameter setting fits the properties of our ISS images in which the reference stars are sparse, generally less than 20 in each image, and displayed as straight lines. It is because all these observations have a small field of view  of $\sim 2’$, a limited magnitude of $\sim 15$, and steadily track the observation target. These three factors result in the reference stars being sparse and appearing as straight lines.
 \par It should be pointed out that not all streak regions detected by our method are true. That is, there are some false streak detections in an ISS image. However, it does not influence the result because  all true streaks will be matched with catalogue stars while the false detections will be filtered out.

The disadvantage of the method is that it cannot  provide the accurate center position for a curved streak. In physical view,  the modified moment method only works out the  first-order moment center of an intensity distribution. For a non-straight streak, the geometric center is not what we desired. It requires further research on centering curved streaks. Fortunately, all trailed stars appear as straight streaks in all images of our targets.
\begin{table}
\bc
\begin{minipage}[]{150mm}
\caption[]{Mean values and standard deviations of residuals of our results relative to the JPL ephemeris SAT427. The others include Dione, Mimas, Rhea and Tethys.\label{table2}}\end{minipage}
\setlength{\tabcolsep}{1pt}
\small
\begin{tabular}{cccccccc}
\hline
                         &      & Sample   & Line     & \(\alpha*\cos{\delta}\) & \(\delta\) & \(\alpha*\cos{\delta}\) & \(\delta\) \\
                         &      & (pixels) & (pixels) & (arcsec)     & (arcsec)  & (km)       & (km)           \\ \hline
Enceladus                & Mean & 0.02     & 0.04     & 0.04         & 0.03      & 0.01       & 0.13           \\
                         & SD  & 0.16     & 0.15     & 0.14         & 0.17      & 0.69       & 0.98           \\
The others                  & Mean & 0.02     & 0.05     & 0.06         & 0.11      & 0.58       & 0.87           \\
                         & SD  & 0.24     & 0.27     & 0.25         & 0.31      & 2.02       & 2.36           \\ 
Total                  & Mean & 0.02     & 0.02     & 0.01         & 0.04      & 0.11       & 0.26           \\
                         & SD  & 0.18     & 0.18     & 0.17         & 0.22      & 1.08       & 1.37           \\ \hline
\end{tabular}
\ec
\end{table}

\section{Conclusion}
\label{sect:section5}
In this paper, a method is proposed to automatically search streaks and determine their accurate centers. The method is applied to measure 658 ISS images of several main moons of Saturn with streaks, 539 of these images are Enceladus, and the remaining images are Mimas, Dione, Tethys and Rhea. The final results show that the proposed method is efficient. Compared with the JPL ephemeris SAT427, mean residuals of all measurements of Enceladus are 0.02 pixels in sample  and 0.04 pixels in line, with standard deviations of 0.16 and 0.15 pixels, respectively. In terms of distance, the mean residuals of Enceladus are 0.01 km in $\alpha \times cos(\delta)$  and 0.13 km in $\delta$, with standard deviation smaller than $1$ km. For the other four targets, the means and standard deviations of their residuals are worse than Enceladus because of the different observation conditions.  Overall, in terms of residuals in linear units, the means in the right  ascension and declination are 0.11 and 0.26 km, respectively. The standard deviations are 1.08 and 1.37 km, respectively.

\normalem
\begin{acknowledgements}
This work has been partly supported by the Joint Research Fund in Astronomy under cooperative agreement between the National Natural Science Foundation of China and Chinese Academy of Sciences ( No. U2031104), and National Natural Science Foundation of China (No. 62071201). This work has made use of data from the European Space Agency (ESA) mission {\it Gaia} (\url{https://www.cosmos.esa.int/gaia}), processed by the {\it Gaia} Data Processing and Analysis Consortium (DPAC, \url{https://www.cosmos.esa.int/web/gaia/dpac/ consortium}). Funding for the DPAC has been provided by national institutions, in particular the institutions participating in the {\it Gaia} Multilateral Agreement.

\end{acknowledgements}
  
\bibliographystyle{raa}
\bibliography{bibtex}

\end{document}